\documentclass[11pt,a4paper]{article}

\usepackage[utf8]{inputenc}

\usepackage[a4paper,margin=1in]{geometry}

\usepackage{amsmath,amssymb}

\usepackage{array}
\usepackage{booktabs}          
\usepackage{multirow}          

\usepackage{graphicx}
\usepackage[dvipsnames,svgnames,x11names]{xcolor}
\usepackage{tikz}
\usetikzlibrary{positioning, arrows.meta, decorations.pathreplacing, shapes.geometric, calc}

\usepackage{algorithm}
\usepackage{algorithmic}

\usepackage{microtype}         
\usepackage{enumitem}          
\setlist{nosep, leftmargin=*}  

\usepackage{hyperref}
\hypersetup{
  colorlinks=true,
  linkcolor=NavyBlue,
  citecolor=OliveGreen,
  urlcolor=BrickRed,
  pdfauthor={AlgoXpert Lab},
  pdftitle={Alpha Research Framework}
}
\usepackage[capitalise,noabbrev]{cleveref}  

\usepackage{authblk}

\definecolor{stageIS}{HTML}{E8F4FD}
\definecolor{stageWFA}{HTML}{FFF3E0}
\definecolor{stageOOS}{HTML}{E8F5E9}
\definecolor{gatecol}{HTML}{FFF9C4}
\definecolor{deploycol}{HTML}{C8E6C9}
\definecolor{rejectcol}{HTML}{FFCDD2}
\definecolor{refactorcol}{HTML}{FFE0B2}
\definecolor{lanecol}{HTML}{F3E5F5}
\definecolor{passcol}{HTML}{2E7D32}
\definecolor{failcol}{HTML}{C62828}

\title{AlgoXpert Alpha Research Framework: A Rigorous IS--WFA--OOS Protocol for Mitigating Overfitting in Quantitative Strategies}

\author[1,4]{Nguyen Thi Nguyet}
\author[2,4]{Nguyen Bao Chan}
\author[3,4]{Pham The Anh\thanks{Founder \& CEO, AlgoXpert}}

\affil[1]{\texttt{ds.nguyetnt@gmail.com}}
\affil[2]{\texttt{nguyenbaochan20042002@gmail.com}}
\affil[3]{\texttt{pta.forwork@gmail.com}}

\affil[4]{AlgoXpert Lab --
Algorithmic Trading, Quantitative Finance, AI \& Data Science\\
\url{https://algoxpert.org/}\\

\textit{Corresponding author:} \texttt{admin@algoxpert.org}}

\date{}

\begin{document}
\maketitle

\begin{abstract}
Transitioning a strategy from backtest to live operation is where many quantitative trading systems fail, primarily due to parameter overfitting, selection bias, and fragility under regime shifts. This paper proposes the \textbf{AlgoXpert Alpha Research Framework}---a standardized, decision-oriented protocol that evaluates strategies across three chronological stages: (i)~In-Sample (IS), which prioritizes stable parameter regions (plateaus) over single optima; (ii)~Walk-Forward Analysis (WFA) with rolling windows and purge gaps to mitigate information leakage, equipped with majority-pass and catastrophic-veto decision gates; and (iii)~Out-of-Sample (OOS) holdout under strict parameter lock (no further tuning). The framework integrates a defense-in-depth architecture: structural (cliff veto), execution (spread/leverage guards), and equity protection (circuit breakers, kill switch). A case study on USDJPY M5 intraday illustrates the procedure for detecting overfitting through performance degradation and drawdown integrity under chronology. A post-validation report comparing four alpha variants (v1--v4) reveals rank reversal when switching the objective from maximizing Sharpe to minimizing MaxDD, underscoring the trade-off between risk-adjusted performance and tail-risk control.

\par\medskip
\noindent\textbf{Keywords:} Alpha Research Framework; Walk-Forward Analysis; Purged Validation; Parameter Stability; Backtest Overfitting; Selection Bias; Execution-Aware Backtesting; Stress Testing; Kill Switch; Out-of-Sample Verification.
\par\medskip
\noindent\textbf{Disclaimer.} For research purposes only; not investment advice. Past performance does not guarantee future results.
\end{abstract}

\section{Introduction}

In algorithmic trading, the gap between a seemingly compelling backtest alpha and a stable live system is where strategies most often fail. Three prevalent causes are: (i) excessive parameter optimization on finite data, causing the model to fit noise (\textit{overfitting})~\cite{bailey14}, (ii) testing numerous configurations or variants without proper controls, leading to selection bias~\cite{harvey16,white00}, and (iii) time-varying changes in volatility, liquidity, and execution costs that render strategies fragile under regime shifts. Accordingly, the objective of this paper is not to maximize ``peak backtest'' performance, but to standardize a \textbf{deployment decision protocol} that is reproducible and auditable: a strategy advances only upon passing clearly defined decision gates in strict chronological order.

\subsection{Practical Gaps}
In practice, many useful techniques exist---walk-forward analysis~\cite{pardo}, parameter stability checks, stress testing, and data-leakage mitigation~\cite{prado18}. However, these are often applied in isolation, lacking an end-to-end \textit{protocol} that translates research results into deployment decisions. This paper targets three \textbf{core gaps} and three \textbf{secondary gaps}:

\begin{itemize}
    \item \textbf{Core Gap A -- Parameter selection without ``stability regions'':} many pipelines still select a single optimum, whereas extreme points are typically sensitive and prone to breakdown under small perturbations; there is a lack of policies that prioritize \textit{stable plateaus} and avoid \textit{cliff}-type sensitivity regions.
    \item \textbf{Core Gap B -- WFA without leakage control for stateful strategies:} na\"ive train--test splitting can yield optimistic evaluations for \textit{stateful/path-dependent} strategies~\cite{prado18}, as indicator overlap and position state may ``bleed'' across boundaries, reducing the ``blindness'' of the forward test.
    \item \textbf{Core Gap C -- Validation decoupled from execution and risk:} many pipelines optimize the \textit{signal} first, appending costs and guards as an afterthought~\cite{kissell14}, whereas operational failures often originate from microstructure friction (spread widening, adverse fills) and ill-timed leverage accumulation.

    \item \textbf{Secondary Gap 1 -- Absence of deployment gates:} explicit pass/fail checklists for the IS$\to$WFA$\to$OOS progression are rare.
    \item \textbf{Secondary Gap 2 -- Safeguards without ablation:} ablation studies that disable individual protection layers to measure their marginal contribution to tail-risk reduction are seldom conducted.
    \item \textbf{Secondary Gap 3 -- Search transparency:} descriptions of \textit{degrees of freedom} (grid/trial sizes) and parameter-locking policies are frequently omitted, impeding assessment of selection bias and reproducibility.
\end{itemize}

\subsection{Proposal and Contributions}
We propose the \textbf{AlgoXpert Alpha Research Framework}, an IS--WFA--OOS pipeline \textit{oriented toward deployment decisions} that aims to reduce overfitting and improve generalizability. Rather than maximizing ``peak backtest'' metrics, the framework prioritizes robustness across multiple validation stages. The main contributions are:

\begin{itemize}
    \item \textbf{(C1) Stability-region parameter selection:}
    instead of selecting a single optimum, we prioritize configurations within a ``stable region'' (plateau) and avoid sensitivity-prone ``cliff'' zones. The IS stability region is defined as:
    \[
        \Omega_{stable}=\{\theta \mid SR(\theta)\ge 0.9\cdot SR_{opt}\},
    \]
    retaining configurations whose risk-adjusted Sharpe achieves at least 90\% of the best observed IS configuration. Details on the search space, $\alpha$ selection, and cliff-veto criteria are provided in Section~\ref{sec:protocol_def}.

    \item \textbf{(C2) Purged rolling walk-forward analysis:}
    we design WFA with rolling windows and insert \textit{purge gaps} between train and test segments to mitigate leakage and carryover effects, which is particularly important for \textit{stateful/path-dependent} strategies (e.g., grid, trailing, inventory state). The protocol description and WFA schedule are detailed in Section~\ref{sec:protocol_def}.

    \item \textbf{(C3) Defense-in-depth safeguards:}
    execution and equity constraints are embedded directly into the pipeline (e.g., spread/leverage guards, circuit breakers, and \textit{kill switch}). Additionally, the framework mandates \textit{stress testing} by degrading execution assumptions (increasing spread/commission or simulating adverse fills) to assess resilience. Details on metrics and stress envelope are provided in Section~\ref{sec:eval_measures}.

    \item \textbf{(C4) Decision gates and checklists:}
    the framework provides pass/fail criteria for the IS$\to$WFA$\to$OOS progression via \textit{decision gates} that are \textbf{pre-committed}. At Stage II, gates are anchored to \textbf{forward windows}: each fold $i$ is deemed \textsc{PASS} only if the metric vector on $W_i^{test}$ meets the minimum benchmark $\mathbf{b}$; the entire WFA is \textsc{PASS} if the proportion of passing folds meets threshold $q$ (majority-pass), and \textsc{FAIL} immediately upon triggering the \textit{catastrophic veto} (e.g., drawdown integrity breakdown or constraint violation $\mathcal{C}$). The majority-pass + catastrophic-veto mechanism and fold-level evaluation procedure are described directly in Algorithm~\ref{alg:wfa}.

    Beyond gate criteria, we additionally report train$\to$test degradation diagnostics (e.g., resilience ratio) to illuminate failure modes; these diagnostics are not used for pass/fail decisions or tuning.

\end{itemize}

\subsection{Scope and Assumptions}
The framework focuses on time-series trading strategies, particularly stateful/path-dependent ones, where information leakage at train--test boundaries and carryover effects readily produce optimistic evaluations. The procedure adheres to chronology with pre-committed decision gates; train$\to$test degradation diagnostics serve failure-mode analysis under a diagnostic-only principle (see Section~\ref{sec:stage_wfa}).

\subsection{Paper Organization}
Section~\ref{sec:related_work} surveys related work. Section~\ref{sec:methods} presents the IS--WFA--OOS protocol, purge/state-normalization mechanisms, and the safeguard architecture. Section~\ref{sec:experiment} illustrates the framework via a case study following chronology. Section~\ref{sec:alpha_comparison} provides a post-validation report for research and portfolio orientation. Section~\ref{sec:conclusion_vn} discusses deployment implications, limitations, and future directions.

\section{Related Work}
\label{sec:related_work}

A strategy may achieve very high Sharpe or returns in backtesting solely because \textit{too many} configurations, variants, or rules have been tested on the same dataset. As the number of trials increases, the probability of a ``lucky hit'' (false discovery) also increases, even when no genuine edge exists. Harvey et al.\ emphasize the risk of false discoveries in expected-return research as the number of hypotheses (factors/strategies) proliferates, necessitating multiple-testing controls \cite{harvey16}. From a practitioner's perspective, Bailey et al.\ characterize backtest overfitting as a natural consequence of R\&D pipelines with large degrees of freedom and performance-based strategy selection \cite{bailey14}. These results reinforce the argument that ``peak backtest'' is insufficient for deployment decisions without a protocol that limits overfitting and selection bias.

Even when a researcher reports only a \textit{single} final strategy, results may still be biased because that strategy was selected after extensive experimentation (data snooping). White proposed the \textit{Reality Check} to test the performance of ``the best strategy in a candidate set'' after data-snooping, rather than testing each strategy independently \cite{white00}. Hansen subsequently introduced the \textit{Superior Predictive Ability (SPA)} test to increase power and reduce sensitivity to low-quality candidates in the comparison set \cite{hansen05}. These lines of work demonstrate that if one's pipeline generates multiple candidates, the evaluation must account for the selection process rather than treating the result as a single-shot test.

A complementary strand adjusts popular metrics to avoid optimistic assessment. Bailey and L\'opez de Prado proposed the \textit{Deflated Sharpe Ratio (DSR)} to correct the observed Sharpe for sample length, non-normality of returns, and selection bias from multiple trials \cite{dsr14}. The pragmatic message is: the ``raw'' Sharpe can be substantially inflated when the strategy is selected from a large search space; therefore one must either (i) reduce degrees of freedom, (ii) employ bias-corrected metrics/tests, or combine both.

WFA is a common practice in systematic strategy evaluation and is typically described as rolling train--forward testing over time \cite{pardo}. However, temporal splitting alone does not guarantee elimination of all forms of leakage. For example, long indicator lookbacks cause overlap at train/test boundaries; or strategies with position management (grid, trailing, inventory) make current decisions dependent on the prior path. In such cases, the forward test may be less ``blind'' than expected, producing optimistic assessments. Therefore, WFA is truly useful for deployment decisions only when accompanied by leakage-control rules and explicit pass/fail gates, rather than serving merely as a performance report.

In finance, especially when labels/features overlap temporally, L\'opez de Prado emphasizes \textit{purging} and \textit{embargo} to reduce information leakage between train and test \cite{prado18}. The intuition is to create a buffer around the cutoff point to mitigate serial correlation and overlap effects. This paper inherits that principle and translates it into a \textit{purge gap} within rolling WFA, targeting both feature-level leakage (indicator overlap) and state-level leakage (state carryover) for path-dependent strategies.

The gap between backtest and live performance often originates from execution friction: spread widening, slippage, thin liquidity, and regime-dependent costs. The algorithmic trading and optimal execution literature demonstrates that observed performance depends heavily on cost/fill assumptions, necessitating stress testing under adverse scenarios (e.g., cost inflation, degraded fill quality) to identify the strategy's ``breaking point'' \cite{kissell14,almgren00}. From a risk-management perspective, circuit breakers and kill switches are operational-level safeguards designed to limit losses during tail events or when market conditions deviate from model assumptions; they function as ``emergency stops'' rather than components of performance optimization.

In summary, the above research strands indicate that: (i) selection bias and multiple testing must be controlled at the process level, (ii) WFA requires leakage-mitigation mechanisms for stateful strategies, and (iii) evaluation must be execution-aware through stress testing and operational safeguards.

In contrast to work that focuses on proposing new alphas, this paper focuses on a deployable workflow: (i) parameter selection via stability regions rather than extreme points, (ii) purged rolling WFA for stateful strategies, (iii) stress testing under execution assumptions with operational safeguards, and (iv) auditable decision gates and checklists. The goal is to reduce selection bias, enhance reproducibility, and improve decision quality when transitioning from research to operations.


\section{Methods}
\label{sec:methods}

\subsection{Overview of the Deployment Decision Protocol (IS$\to$WFA$\to$OOS)}
\label{sec:strategy_model}

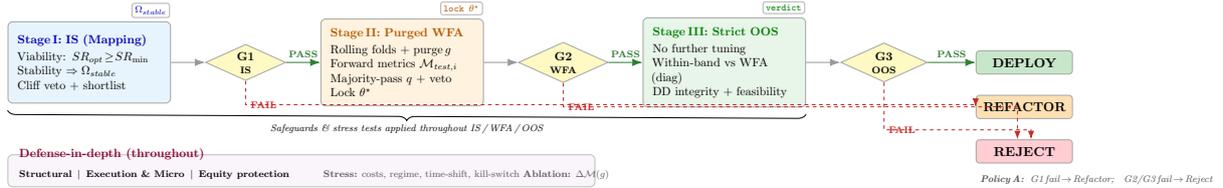
\begin{figure}[!t]
\centering
\resizebox{\linewidth}{!}{%
\begin{tikzpicture}[
  font=\small,
  node distance=10mm and 10mm,
  stage/.style={
    draw=gray!70, rounded corners=4pt, align=left, inner sep=8pt,
    text width=4.2cm, minimum height=2.4cm,
    line width=0.8pt
  },
  gate/.style={
    draw=gray!70, diamond, aspect=2.2, inner sep=2pt, align=center,
    fill=gatecol, line width=0.8pt, font=\small\bfseries
  },
  outcome/.style={
    draw=gray!70, rounded corners=3pt, align=center, inner sep=6pt,
    text width=2.4cm, line width=0.8pt, font=\bfseries
  },
  lane/.style={
    draw=gray!50, rounded corners=4pt, align=left, inner sep=10pt,
    fill=lanecol!40, line width=0.6pt, text width=16.5cm
  },
  tag/.style={
    draw=gray!60, rounded corners=2pt, inner sep=2.5pt,
    font=\ttfamily\scriptsize, align=center, fill=white
  },
  arrow/.style={-{Latex[length=3mm]}, thick, color=gray!80},
  passarrow/.style={-{Latex[length=3mm]}, thick, color=passcol},
  fail/.style={-{Latex[length=2.5mm]}, thick, dashed, color=failcol}
]

\node[lane] (lane) {\vphantom{X}\\[1mm]};
\node[anchor=west, font=\bfseries\small, color=purple!70!black]
  at ([xshift=6pt]lane.north west) {Defense-in-depth (throughout)};
\node[anchor=west, font=\scriptsize] at ([xshift=6pt,yshift=-17pt]lane.north west)
  {\textbf{Structural}\;\;$\vert$\;\;\textbf{Execution \& Micro}\;\;$\vert$\;\;\textbf{Equity protection}
   \qquad\quad
   \textcolor{gray!70!black}{\textbf{Stress:} costs, regime, time-shift, kill-switch}
   \hfill
   \textcolor{gray!70!black}{\textbf{Ablation:} $\Delta\mathcal{M}(g)$}};

\node[stage, fill=stageIS, above=32mm of lane.west, anchor=west] (is) {
  \textbf{\textcolor{blue!70!black}{Stage\,I: IS (Mapping)}}\\[2pt]
  Viability: $SR_{opt}\!\ge\! SR_{\min}$\\
  Stability $\Rightarrow \Omega_{stable}$\\
  Cliff veto + shortlist
};

\node[gate, right=of is] (g1) {G1\\[-1pt]{\scriptsize IS}};

\node[stage, fill=stageWFA, right=of g1] (wfa) {
  \textbf{\textcolor{orange!70!black}{Stage\,II: Purged WFA}}\\[2pt]
  Rolling folds + purge\,$g$\\
  Forward metrics $\mathcal{M}_{test,i}$\\
  Majority-pass $q$ + veto\\
  Lock $\theta^\star$
};

\node[gate, right=of wfa] (g2) {G2\\[-1pt]{\scriptsize WFA}};

\node[stage, fill=stageOOS, right=of g2] (oos) {
  \textbf{\textcolor{green!50!black}{Stage\,III: Strict OOS}}\\[2pt]
  No further tuning\\
  Within-band vs WFA (diag)\\
  DD integrity + feasibility
};

\node[gate, right=of oos] (g3) {G3\\[-1pt]{\scriptsize OOS}};

\node[tag, anchor=south east] at ([xshift=-1pt,yshift=3pt]is.north east)
  {\textcolor{blue!70!black}{$\Omega_{stable}$}};
\node[tag, anchor=south east] at ([xshift=-1pt,yshift=3pt]wfa.north east)
  {\textcolor{orange!70!black}{lock $\theta^\star$}};
\node[tag, anchor=south east] at ([xshift=-1pt,yshift=3pt]oos.north east)
  {\textcolor{green!50!black}{verdict}};

\node[outcome, fill=deploycol, right=14mm of g3]  (deploy)   {DEPLOY};
\node[outcome, fill=refactorcol, below=6mm of deploy]  (refactor) {REFACTOR};
\node[outcome, fill=rejectcol, below=6mm of refactor] (reject)   {REJECT};

\node[anchor=north west, font=\scriptsize\itshape, text=gray!60!black]
  at ([xshift=0mm,yshift=-2mm]reject.south west)
  {\textbf{Policy\,A:}\; G1\,fail\,$\to$\,Refactor;\;\;
   G2/G3\,fail\,$\to$\,Reject};

\draw[arrow] (is.east) -- (g1.west);
\draw[passarrow] (g1.east) -- node[above, font=\scriptsize\bfseries, color=passcol]{PASS} (wfa.west);
\draw[arrow] (wfa.east) -- (g2.west);
\draw[passarrow] (g2.east) -- node[above, font=\scriptsize\bfseries, color=passcol]{PASS} (oos.west);
\draw[arrow] (oos.east) -- (g3.west);
\draw[passarrow] (g3.east) -- node[above, font=\scriptsize\bfseries, color=passcol]{PASS} (deploy.west);

\draw[fail] (g1.south) -- ++(0,-7mm)
  -| node[pos=0.0, right, font=\scriptsize\bfseries, color=failcol]{FAIL}
  (refactor.west);

\draw[fail] (g2.south) -- ++(0,-7mm)
  -| node[pos=0.0, right, font=\scriptsize\bfseries, color=failcol]{FAIL}
  ([xshift=-2mm]reject.north);

\draw[fail] (g3.south) -- ++(0,-14mm)
  -| node[pos=0.0, right, font=\scriptsize\bfseries, color=failcol]{FAIL}
  ([xshift=2mm]reject.north);

\draw [thick, decorate, decoration={brace, amplitude=7pt, mirror}]
  ([yshift=-2mm]is.south west) -- ([yshift=-2mm]oos.south east)
  node[midway, yshift=-14pt, font=\scriptsize\itshape]
  {Safeguards \& stress tests applied throughout IS\,/\,WFA\,/\,OOS};

\end{tikzpicture}%
}
\caption{IS$\to$WFA$\to$OOS deployment decision protocol with decision gates.}
\label{fig:protocol_overview}
\end{figure}

The goal of this work is to standardize a \textbf{deployment decision protocol} that reduces overfitting and selection bias in quantitative strategy R\&D. The protocol enforces \textbf{chronology} and \textbf{decision gates} of the pass/fail type: a strategy advances only upon passing validation in strict chronological order.

As illustrated in Figure~\ref{fig:protocol_overview}, the framework comprises three stages:
(i) \textbf{IS} performs stability mapping to identify $\Omega_{stable}$ (plateau) and discard cliff-sensitive configurations;
(ii) \textbf{purged rolling WFA} evaluates sequential adaptability through rolling folds with purge gap $g$, accompanied by diagnostics (e.g., $\eta_i$) and gate/veto mechanisms to lock the final configuration $\theta^\star$;
(iii) \textbf{strict OOS holdout} provides the final validation under \textbf{no further tuning}. Pass/fail thresholds are \textbf{pre-committed} before opening OOS.

In parallel with the three stages, the protocol integrates \textbf{defense-in-depth safeguards} and a \textbf{stress-test battery} (execution/microstructure, equity protection), and defines ablation $\Delta\mathcal{M}(g)$ to measure the marginal contribution of each safeguard.

\paragraph{Decision states.}
The protocol returns one of three outcomes: \textbf{Deploy} (all gates passed), \textbf{Reject} (fail/veto), or \textbf{Refactor} (failure mode is structural in nature, requiring design changes rather than parameter tuning).

\paragraph{Branching policy (Policy~A).}
In this paper, we adopt \textbf{Policy A} for branching decisions:
\emph{G1 fail $\Rightarrow$ Refactor} (structural/design-level failure),
\emph{G2 fail $\Rightarrow$ Reject} (insufficient sequential adaptability), and
\emph{G3 fail $\Rightarrow$ Reject} (failed strict holdout).

\subsection{Strategy Class and Notation}
\label{sec:protocol_def}

Consider a stateful, path-dependent algorithmic trading strategy, applicable to position-management paradigms such as trailing/grid/inventory. Market data are denoted $D=\{x_t\}_{t=1}^T$, where $x_t$ may include bid/ask prices, spread, and other necessary microstructure variables. The strategy is parameterized by a vector $\theta\in\Theta$ and generates a sequence of trading decisions over time.

\paragraph{State and action.}
At time $t$, the strategy maintains an internal state $s_t$ (e.g., open positions, grid level, trailing state, margin usage). Based on the observation $x_t$ and state $s_t$, the strategy produces an action $a_t\in\mathcal{A}$ (open/close/adjust position, choose lot size, set/adjust stop/take-profit), while simultaneously updating the state:
\begin{equation}
a_t=\pi_\theta(x_t,s_t),\qquad s_{t+1}=f_\theta(s_t,x_t,a_t),
\end{equation}
where $\pi_\theta$ is the trading policy and $f_\theta$ describes the state dynamics. Making $s_t$ explicit is important because, for path-dependent strategies, na\"ive temporal train/test splitting can introduce leakage at window boundaries (due to indicator/lookback overlap and position state ``bleeding'' across boundaries), making the forward test less ``blind'' than expected. This directly motivates the purged WFA design in Section~\ref{sec:strategy_model}.

\paragraph{Execution and risk constraints.}
Let $\mathcal{C}$ denote the set of execution/risk constraints (e.g., spread guard, leverage/DL limit, position count limit, circuit breakers/kill-switch). A parameter configuration $\theta$ is considered feasible if, throughout the entire simulation, all generated actions satisfy these constraints:
\begin{equation}
\pi_\theta \text{ feasible} \iff (x_t,s_t,a_t)\models \mathcal{C}\quad \forall t.
\end{equation}
In our framework, $\mathcal{C}$ is enforced throughout IS/WFA/OOS and during stress tests, ensuring that the validation process is execution-aware rather than purely signal-optimized.

\paragraph{Equity curve and risk measures.}
Let $E_t$ denote equity at time $t$. The running peak through time $t$ is
$E_{\mathrm{peak}}(t)=\max_{u\le t}E_u$.
Equity-based drawdown at time $t$ is defined as:
\begin{equation}
\mathrm{DD}_{eq}(t)=\frac{E_{\mathrm{peak}}(t)-E_t}{E_{\mathrm{peak}}(t)},\qquad
\mathrm{MDD}_{eq}=\max_t \mathrm{DD}_{eq}(t).
\end{equation}
Performance and robustness measures (e.g., Sharpe, Calmar, recovery factor, worst-window loss, turnover/holding-time proxies) and their formal estimation/reporting procedures are presented in Section~\ref{sec:eval_measures}. In the methods section, the above notation is used to describe the decision gates, particularly the ``drawdown integrity'' criterion in the OOS stage and the equity-protection layers in the safeguards.

\subsection{Stage~I: In-Sample Stability Mapping}
\label{sec:stage_is}

The IS stage aims to produce a set of \textbf{robust} parameter candidates for WFA (Section~\ref{sec:stage_wfa}), rather than selecting a single backtest optimum~\cite{bailey14}. Extreme points are typically sensitive to data noise, regime changes, and minor execution-assumption discrepancies; accordingly, the framework prioritizes ``plateaus'' (near-optimal but stable regions) and discards ``cliffs'' (sensitive zones where small parameter perturbations can cause performance collapse or drawdown spikes).

\subsubsection{Stability Region $\Omega_{stable}$ (Plateau Selection)}
On the IS dataset, we perform a performance mapping over a finite search space (grid or trials with budget $B$) to observe the topology of $SR(\theta)$, while simultaneously tracking minimum tail-risk indicators (particularly $\mathrm{MDD}_{eq}(\theta)$). Let
\begin{equation}
SR_{opt}=\max_{\theta\in\Theta} SR(\theta)
\end{equation}
denote the best \emph{observed} value in IS under budget $B$ (not to be interpreted as the true optimum).

\paragraph{Viability condition.}
Stage I is considered viable only if
\begin{equation}
SR_{opt} \ge SR_{\min} > 0.
\end{equation}
If this condition is not met, the strategy terminates with verdict \textbf{Refactor} (Policy A), rather than proceeding to plateau selection.

The stability region is defined as:
\begin{equation}
\Omega_{stable}=\{\theta\in\Theta \mid SR(\theta)\ge \alpha \cdot SR_{opt}\},
\end{equation}
where $\alpha\in(0,1)$ is the plateau threshold (default $\alpha=0.9$; sensitivity analysis with respect to $\alpha$ is reported in Section~\ref{sec:analysis}). Configurations outside $\Omega_{stable}$ are excluded due to markedly inferior performance or the risk of spurious selection under large degrees of freedom.

\paragraph{Trade-count feasibility filter.}
To avoid configurations with high $SR$ but based on too few observations (unreliable and potentially lucky), we apply a minimum trade-count threshold in IS:
\begin{equation}
N_{\text{trades}}(\theta)\ge N_{\min}.
\end{equation}
Configurations failing this threshold are removed from $\Omega_{stable}$ before shortlisting.

\paragraph{Ranking within $\Omega_{stable}$.}
After plateau and feasibility filtering, ranking/shortlisting is performed \textbf{within} $\Omega_{stable}$. Rather than maximizing $SR$ (which risks reverting to extreme-point selection), we prioritize criteria related to tail-risk and recovery (e.g., Calmar, recovery factor, worst-window loss), aiming to select candidates with reasonable risk characteristics before advancing to WFA.

\subsubsection{Cliff Filtering (Sensitivity Veto)}
The objective of cliff filtering is to remove configurations that are ``good but thin'': a very small parameter change causes sharp performance degradation or drawdown spikes. For each $\theta$, consider the neighborhood $\mathcal{N}(\theta)$ consisting of adjacent points along each parameter dimension with displacement equal to one grid resolution step (e.g., $\pm 1$ step), holding other dimensions fixed.

\paragraph{Performance cliff (Sharpe).}
\begin{equation}
\mathrm{Cliff}_{SR}(\theta) = \max_{\theta'\in \mathcal{N}(\theta)} \left[ SR(\theta)-SR(\theta') \right]_+,
\end{equation}
where $[z]_+=\max(z,0)$ measures the worst-case Sharpe degradation under minimal perturbation.

\paragraph{Tail-risk cliff (equity drawdown).}
\begin{equation}
\mathrm{Cliff}_{DD}(\theta) = \max_{\theta'\in \mathcal{N}(\theta)} \left[ \mathrm{MDD}_{eq}(\theta')-\mathrm{MDD}_{eq}(\theta) \right]_+.
\end{equation}

\paragraph{Rejection rule.}
A configuration is rejected if it is sensitive under either criterion:
\begin{equation}
\theta \text{ is rejected if }
\mathrm{Cliff}_{SR}(\theta)>\tau_{SR}
\ \ \text{or}\ \
\mathrm{Cliff}_{DD}(\theta)>\tau_{DD}.
\end{equation}
In practice, cliff filtering is applied \textbf{within} $\Omega_{stable}$ to favor plateau regions with ``thickness'' (robustness) over thin peaks (fragility).

\subsubsection{Parameter Locking Policy}
After Stage I, the framework applies a \textbf{parameter locking} policy to constrain degrees of freedom and reduce selection bias across subsequent time windows:

\begin{itemize}
    \item \textbf{Lock time:} locking occurs after completing plateau selection, feasibility filtering, and cliff veto, simultaneously finalizing a finite shortlist.
    \item \textbf{Locked dimensions:} cliff-prone parameter dimensions or those governing high-risk behavior (e.g., core parameters or those increasing position/leverage accumulation) are fixed, or restricted to a very small pre-reported subset.
    \item \textbf{WFA degrees of freedom:} during WFA, each fold may only re-select $\theta_i$ from the shortlist (or the locked subset), without re-opening the entire $\Theta$ or altering the strategy structure.
\end{itemize}

The shortlist size, search budget $B$, $\alpha$ value, threshold $N_{\min}$, and cliff thresholds $(\tau_{SR},\tau_{DD})$ must be reported to allow readers to assess the degrees of freedom, selection bias, and reproducibility of the pipeline.

\subsection{Stage~II: Purged Rolling Walk-Forward Analysis}
\label{sec:stage_wfa}

The WFA stage~\cite{pardo} is designed to test the \textbf{sequential adaptability} of the strategy: an alpha is considered \textit{deployable} only if it maintains quality when shifted through time. For \textit{stateful/path-dependent} strategies (grid/trailing/inventory), na\"ive train/test splitting can produce optimistic evaluations due to (i) lookback overlap at window boundaries, and (ii) \textit{state carryover}~\cite{prado18}. Accordingly, we employ \textbf{purged rolling WFA} and impose pass/fail \textbf{decision gates} before opening OOS.

\paragraph{Procedure summary.}
Algorithm~\ref{alg:wfa} describes the WFA in pseudocode, emphasizing restricted DoF, purge, state normalization, majority-pass gate, and catastrophic veto.

\begin{algorithm}[!t]
\caption{Purged rolling WFA with majority-pass gate and catastrophic veto.}
\label{alg:wfa}
\begin{algorithmic}[1]
\REQUIRE Stability region/shortlist $\Omega_{stable}$; schedule $\{(W_i^{train},g,W_i^{test})\}_{i=1}^{N}$; constraints $\mathcal{C}$;
forward benchmarks $\mathbf{b}$; majority threshold $q$; catastrophic veto rule.
\ENSURE WFA verdict \textsc{PASS}/\textsc{FAIL} and locked parameters $\theta^\star$ if \textsc{PASS}.

\STATE $pass \leftarrow 0$, $\mathcal{I}\leftarrow \emptyset$.
\FOR{$i=1$ to $N$}
  \STATE \textbf{Optimize (restricted DoF):} choose $\theta_i \in \Omega_{stable}$ using only $W_i^{train}$.
  \STATE \textbf{Purge:} exclude the gap of length $g$ immediately after $W_i^{train}$.
  \STATE \textbf{State normalization:} reset strategy state at start of $W_i^{test}$ (e.g., flat/no inventory).
  \STATE \textbf{Forward test:} run $\pi_{\theta_i}$ on $W_i^{test}$ under $\mathcal{C}$; compute $\mathcal{M}_{test,i}$.
  \IF{fold $i$ is not evaluable (data/simulation failure or insufficient forward sample)}
     \STATE \textbf{continue}
  \ENDIF
  \STATE $\mathcal{I}\leftarrow \mathcal{I}\cup\{i\}$.
  \IF{catastrophic veto triggers on $W_i^{test}$}
     \STATE \textbf{return} \textsc{FAIL}.
  \ENDIF
  \IF{$\mathcal{M}_{test,i}\succeq \mathbf{b}$}
     \STATE $pass \leftarrow pass+1$.
  \ENDIF
\ENDFOR
\IF{$|\mathcal{I}|=0$}
  \STATE \textbf{return} \textsc{FAIL}.
\ENDIF
\IF{$pass/|\mathcal{I}| \ge q$}
  \STATE \textbf{Lock:} select and lock $\theta^\star$ by a pre-committed rule; \textbf{return} \textsc{PASS}.
\ELSE
  \STATE \textbf{return} \textsc{FAIL}.
\ENDIF
\end{algorithmic}
\end{algorithm}

\subsubsection{Rolling Fold Structure and Purge Gap $g$}
The WFA period is divided into $N$ folds according to a fixed, pre-defined schedule. Each fold $i$ consists of a training window, followed by a purge gap, and then an out-of-sample test window:
\begin{equation}
W_i^{train}\ \to\ \text{purge gap } g\ \to\ W_i^{test}.
\end{equation}
The purge gap $g$ is a time segment excluded immediately after $W_i^{train}$ to mitigate overlap/serial-dependence effects between train and test (especially when features have long lookbacks), and to reduce leakage risk from stateful strategies. For intraday data, $g$ is expressed in trading days or bars and is \textbf{fixed} across all folds to avoid result-dependent tuning.

\paragraph{Note on rolling schedule and purge/train overlap between folds.}
In a consecutive rolling design, the purge gap of the preceding fold may overlap with the beginning of $W_{i+1}^{train}$. This \textbf{does not compromise} the ``blindness'' of the forward test, since the critical condition to maintain is: \emph{each} $W_i^{test}$ is always separated from \emph{its own} $W_i^{train}$ by a purge gap immediately preceding it. In other words, the purge is defined by the \emph{intra-fold train--test relationship}, not by the inter-fold distance.

\subsubsection{State Normalization Prior to Forward Testing}
To control leakage from \textit{state carryover} in path-dependent strategies, the forward test of each fold is evaluated under a \textbf{normalized state} assumption at the start of $W_i^{test}$. Specifically, at the test-window boundary, the strategy is reset to a canonical state (e.g., \textit{flat/no inventory}, grid level = 0, trailing state reset, all internal counters/state variables reinitialized).

In terms of implementation, state normalization can be realized in one of two equivalent ways:
(i) run the $W_i^{test}$ simulation as an independent session starting exactly at the test-window opening; or
(ii) force-close all positions and reinitialize internal state variables immediately before entering $W_i^{test}$.
The objective is to ensure that performance within $W_i^{test}$ reflects temporal generalizability, rather than benefiting from accumulated state in prior segments.

\subsubsection{Re-Optimization Rule and Degrees-of-Freedom Constraints}
Within each fold $i$, configuration selection is permitted only within the controlled scope from Stage I:
\begin{equation}
\theta_i \in \Omega_{stable}\quad \text{(or a locked shortlist from Stage I)}.
\end{equation}
This rule ensures WFA does not degenerate into unconstrained optimization across multiple windows (which would inflate selection bias). Per the protocol, WFA is \textbf{not} permitted to (i) re-open the entire parameter space $\Theta$, or (ii) alter the strategy structure. All permitted ``degrees of freedom'' (e.g., selection restricted to a shortlist of size $K$; or only a subset of execution parameters may be adjusted) must be stated explicitly and held fixed across all folds.

Upon completing WFA, a final configuration $\theta^\star$ is selected according to a \textbf{pre-committed} rule, for example: the configuration with the best median-$SR_{test,i}$ across valid folds, or selection based on risk criteria (Calmar/MDD) across test windows. Crucially, the rule for selecting $\theta^\star$ must be finalized before opening OOS.

\subsubsection{Per-Fold Forward Metrics and Gate Design Principles}
For each fold $i$, denote the Sharpe ratio on train and forward test:
\begin{equation}
SR_{train,i} := SR(\pi_{\theta_i}; W_i^{train}),\qquad
SR_{test,i}  := SR(\pi_{\theta_i}; W_i^{test}),
\end{equation}
where $SR(\cdot)$ is defined and reported in detail in Section~\ref{sec:eval_measures}. On the forward window, we simultaneously evaluate risk/feasibility metrics (e.g., $\mathrm{MDD}_{eq,test,i}$, $\mathrm{Calmar}_{test,i}$, trade density, and the degree of $\mathcal{C}$ violation).

\paragraph{Principles.}
\textbf{Decision gates are anchored to forward windows.} Rationale: $W_i^{test}$ constitutes the temporally ``blind'' test, and therefore the WFA pass/fail criteria are constructed primarily from metrics on $W_i^{test}$ (benchmark-based).
\textbf{Train metrics serve only operational and diagnostic roles.} Metrics on $W_i^{train}$ primarily support $\theta_i$ selection within restricted DoF; if needed, train$\to$test indicators may be reported as diagnostics, but they do not constitute a tuning loop and are not primary gate conditions.

\subsubsection{WFA Decision Gate: Majority-Pass + Catastrophic Veto (Pre-committed)}
Let $\mathbf{b}$ denote the minimum benchmark vector on the forward window (e.g., comprising Sharpe, drawdown, Calmar thresholds, and trade-density conditions), and $q\in(0,1]$ the minimum proportion of folds that must meet the benchmark (pre-committed). A fold $i$ is deemed \textbf{PASS} on the forward window if:
\begin{equation}
\mathcal{M}_{test,i}\ \succeq\ \mathbf{b},
\end{equation}
where $\mathcal{M}_{test,i}$ is the metric vector computed on $W_i^{test}$, and $\succeq$ denotes ``meets all thresholds.''

\paragraph{(1) Majority-pass persistence.}
Let $\mathcal{I}$ denote the set of \textit{evaluable} folds (no data/simulation failures and meeting minimum evaluation conditions). The requirement is:
\begin{equation}
\frac{1}{|\mathcal{I}|}\sum_{i\in\mathcal{I}}\mathbb{I}\big[\mathcal{M}_{test,i}\succeq \mathbf{b}\big]\ \ge\ q.
\end{equation}

\paragraph{(2) Catastrophic veto.}
If any $W_i^{test}$ exhibits a catastrophic violation according to the pre-committed veto rule (e.g., drawdown exceeding the integrity threshold, or kill-switch/leverage-cap violation per $\mathcal{C}$), the entire WFA \textbf{FAIL}s immediately.

\paragraph{(3) OOS is not opened upon WFA FAIL.}
Following the decision-oriented principle, if WFA \textbf{FAIL}s, the strategy does not proceed to OOS holdout (verdict: \textbf{Reject} or \textbf{Refactor} depending on failure mode). If WFA \textbf{PASS}es, the final parameters $\theta^\star$ are \textbf{locked} and the strategy advances to Stage III under \textbf{no further tuning}.

\paragraph{Train$\to$test diagnostics (not used for gating).}
\label{par:diag_principle}
Beyond gate criteria, we additionally report $\Delta SR_i := SR_{test,i}-SR_{train,i}$ and resilience ratio $\eta_i := SR_{test,i}/SR_{train,i}$ for failure-mode analysis. \textbf{Overarching principle:} diagnostics serve only post-hoc analysis and are \textbf{not} used as gate conditions or for tuning $\theta^\star$.


\section{Empirical Study}
\label{sec:experiment}

\subsection{Experimental Design}
\label{sec:exp_setting}

The objective of the experimental design is to \emph{operationalize} the IS$\to$WFA$\to$OOS protocol described in Section~\ref{sec:methods} on a minimal but auditable case study that respects \textbf{chronology} (no peeking) and introduces \textbf{no additional degrees of freedom} beyond what the protocol permits.

\subsubsection{Data, Market, and Execution Assumptions}
\label{sec:dataset_exec}

\paragraph{Market and frequency.}
The case study uses the USDJPY pair on the M5 timeframe. Backtests are run using the \textit{Every Tick} simulation mode of MetaTrader 5, reflecting the tester's tick-driven order-matching mechanism.

\paragraph{Data source and time span.}
Data are sourced from broker Exness; the backtest period covers 01/01/2022--31/12/2025. Throughout the experiment, data are used in chronological order, and all decisions/estimates at each stage rely only on the data permitted for observation at that stage.

\paragraph{Capital and leverage assumptions.}
The initial deposit is 100{,}000 with nominal leverage of 1:100. These settings are held fixed throughout IS/WFA/OOS to avoid confounding environmental changes with alpha quality changes.

\paragraph{Execution model and transaction costs.}
The baseline setup uses an \textit{ideal execution} model per the tester (no latency simulation; no adverse slippage beyond the default tick-driven mechanism). Spread/commission assumptions (if separately configured) are considered part of the backtest environment and must be documented under ``Backtest Assumptions'' to ensure reproducibility and proper inference scope. Results in the experimental section should therefore be interpreted as a \textbf{baseline under simplified execution}.

\paragraph{Execution-awareness and stress envelope (reporting scope).}
Per the logic of Gap C, a complete deployment evaluation requires a \textit{stress-test battery} (e.g., increased spread/commission, simulated adverse fills). However, in the current reporting version, the team has \textbf{not} conducted stress branches due to resource constraints; instead, we describe the stress envelope as a \textbf{mandatory framework template} and defer stress/ablation results to future versions (see discussion in Section~\ref{sec:conclusion_vn}).

\subsubsection{IS/WFA/OOS Splits and WFA Schedule (Rolling + Purge)}
\label{sec:splits_wfa_schedule}

\paragraph{Time-ordered splits (no peeking).}
The timeline is partitioned in strict chronological order to reflect deployment conditions:
\begin{itemize}
  \item \textbf{In-Sample (IS):} 01/01/2022--31/12/2023,
  \item \textbf{Walk-Forward Analysis (WFA):} 01/01/2024--31/12/2024,
  \item \textbf{Out-of-Sample holdout (OOS):} 01/01/2025--31/12/2025.
\end{itemize}
The operating principle is: after completing WFA and locking the final parameters $\theta^\star$, \textbf{no further optimization or tuning} is performed when opening OOS (strict no-tuning).

\paragraph{Rolling WFA with fixed purge gap.}
WFA is designed with rolling windows comprising 3 folds; each fold has the structure
$W_i^{train}\to g\to W_i^{test}$ with a fixed purge gap $g=5$ trading days to mitigate leakage from overlap/lookback and carryover effects in stateful strategies. The WFA schedule is \textbf{pre-defined} and held constant throughout the report (Table~\ref{tab:wfa_schedule_vn}).

\begin{table}[!t]
\centering
\caption{Rolling 3-fold WFA schedule ($g=5$ trading days).}
\label{tab:wfa_schedule_vn}
\begin{tabular}{cccc}
\toprule
Fold & $W_i^{train}$ & Purge $g$ & $W_i^{test}$ \\
\midrule
1 & 01/01/2024\,--\,31/03/2024 & 01/04\,--\,05/04 & 06/04\,--\,30/06/2024 \\
2 & 01/04/2024\,--\,30/06/2024 & 01/07\,--\,05/07 & 06/07\,--\,30/09/2024 \\
3 & 01/07/2024\,--\,30/09/2024 & 01/10\,--\,05/10 & 06/10/2024\,--\,31/12/2024 \\
\bottomrule
\end{tabular}
\end{table}

\paragraph{Stateful/path-dependent strategies and state normalization.}
Consistent with the strategy definition in Section~\ref{sec:protocol_def}, each forward window $W_i^{test}$ is evaluated with a \textbf{normalized initial state} at the start (e.g., flat/no inventory) to mitigate \textit{state carryover}. This convention is a necessary condition for interpreting WFA as a temporal validation rather than a sequential concatenation susceptible to state ``bleed-through.''

\paragraph{Degrees-of-freedom constraints in WFA.}
Within each fold, parameter re-selection is permitted only within the controlled scope from Stage~I (restricted DoF; shortlist/$\Omega_{stable}$), and the $\theta^\star$ locking rule after WFA is pre-committed. Train$\to$test diagnostics follow the principle stated in Section~\ref{sec:stage_wfa}.

\subsection{Evaluation Metrics and Decision Gates}
\label{sec:eval_measures}

\subsubsection{Notation and Return Conventions}
Let $E_t$ denote \textit{equity} at time $t$ (reflecting P\&L under the bid/ask model and cost configuration of the backtest).
Throughout IS/WFA/OOS, we use a consistent return frequency to avoid discrepancies when comparing across stages.
Specifically, let the periodic return series be:
\begin{equation}
R_t=\frac{E_t-E_{t-1}}{E_{t-1}}, \qquad t=1,\dots,T,
\end{equation}
where ``period'' may be \textit{daily} (recommended for reporting, consistent with standard statistics) or per-bar/per-trade (if the internal reporting system uses that convention). When annualizing, use the factor $K$ equal to the number of periods per year under the chosen convention (e.g., $K=252$ for daily returns).

\subsubsection{Primary Performance and Risk Metrics}
\paragraph{CAGR (Compound Annual Growth Rate).}
Let $E_{start}$ and $E_{end}$ denote the equity at the beginning and end of the evaluation period; $n$ is the period length in years (under the same convention as $K$):
\begin{equation}
\mathrm{CAGR}=\left(\frac{E_{end}}{E_{start}}\right)^{\frac{1}{n}}-1.
\end{equation}

\paragraph{Max drawdown (equity-based).}
Let $E_{\mathrm{peak}}(t)=\max_{u\le t}E_u$, then:
\begin{equation}
\mathrm{DD}_{eq}(t)=\frac{E_{\mathrm{peak}}(t)-E_t}{E_{\mathrm{peak}}(t)},\qquad
\mathrm{MDD}_{eq}=\max_t \mathrm{DD}_{eq}(t).
\end{equation}

\paragraph{Return series from equity (equity-based).}
Let $E_t$ denote equity (including unrealized P\&L) at sampling point $t$ (e.g., daily or per M5 bar).
The return series is defined from equity:
\begin{equation}
R_t = \ln\left(\frac{E_t}{E_{t-1}}\right)
\quad \text{(or } R_t=\frac{E_t-E_{t-1}}{E_{t-1}}\text{).}
\end{equation}

\textbf{Sharpe ratio (equity-based)~\cite{sharpe66}.}
Given $\{R_t\}_{t=1}^{T}$ and $\sigma_R$ the standard deviation of $R_t$:
\begin{equation}
SR=\frac{\overline{R}-r_f}{\sigma_R}.
\end{equation}
If annualized, $SR_{\mathrm{ann}}=SR\cdot\sqrt{K}$, where $K$ is the number of sampling periods per year
(e.g., $K=252$ if $R_t$ is daily returns).

\paragraph{Calmar ratio.}
\begin{equation}
\mathrm{Calmar}=\frac{\mathrm{CAGR}}{\mathrm{MDD}_{eq}}.
\end{equation}

\subsubsection{Feasibility and Execution-Awareness}
\paragraph{Trade density and sample thickness.}
Let $N_{\mathrm{trades}}$ denote the number of trades (or round-turns per reporting convention) in the period; in addition to the total, report trades/day (over actual trading days) to flag cases of ``good performance from infrequent trading.''

\paragraph{Break-even cost proxy per trade.}
A pragmatic proxy for the maximum average ``cost cushion'' the strategy can absorb before breaking even:
\begin{equation}
C_{max}=\frac{\sum_i P_i}{N_{\mathrm{trades}}}.
\end{equation}
If $C_{max}$ is less than the observed/assumed execution cost (spread+commission+slippage), the strategy is flagged as \textit{execution-fragile}.

\subsubsection{Minimum Benchmarks and Decision Gates (Pre-committed)}
Let $\mathbf{b}$ denote the minimum benchmark vector for PASS/FAIL (committed before opening OOS). In this case study, we use:
\[
SR \ge 2.0,\quad \mathrm{Calmar}\ge 1.5,\quad \mathrm{MDD}_{eq}<7\%,\quad \text{trades/day}\ge 5.
\]
When reporting, if ``trades/day'' cannot be computed directly from logs, this should be noted explicitly and treated as an \textit{open item} rather than assumed to pass.

\paragraph{Stage I (IS) --- viability + shortlist.}
IS \textsc{PASS} if $\mathbf{b}$ is met \textbf{and} $N_{\mathrm{trades}}\ge N_{\min}$ (minimum threshold to guard against thin-sample luck). Only then are plateau selection and cliff veto applied to produce the shortlist.

\paragraph{Stage II (WFA).}
Fold-pass if metrics on $W_i^{test}$ meet $\mathbf{b}$; overall WFA follows majority-pass + catastrophic veto (Section~\ref{sec:stage_wfa}) with $q=2/3$.

\paragraph{Stage III (OOS holdout).}
Parameters are locked after WFA; verdict \textsc{PASS} if OOS metrics meet $\mathbf{b}$ under the same baseline execution.

\subsubsection{Degradation Diagnostics}
Resilience ratio $\eta_i$ and $\Delta SR_i$ (defined in Section~\ref{sec:stage_wfa}) are reported following the diagnostic-only principle stated therein.

\subsection{Experimental Results}
\label{sec:exp_results}

\subsubsection{Stage I (IS): Benchmark Check and Viability}
Over the IS period (01/01/2022--31/12/2023), the strategy achieved:
Sharpe $=2.12$, Calmar $\approx 1.69$, MaxDD (equity) $=6.46\%$, $N=2625$ trades.
Per the benchmark vector $\mathbf{b}$ (Section~\ref{sec:eval_measures}), the Sharpe/Calmar/MaxDD thresholds are met.
The trade-density criterion (trades/day) has not been directly reported; this is noted as a \emph{reporting gap} to be addressed.

\subsubsection{Stage II (WFA): Forward-Window Results}
Table~\ref{tab:wfa_fold_results} summarizes metrics on $W_i^{test}$ and gate status per pre-committed benchmarks. Train metrics serve internal audit only (diagnostic-only principle, Section~\ref{sec:stage_wfa}).

Aggregate WFA (mean across forward segments):
Sharpe $=3.79$, MaxDD $=2.93\%$, Calmar $=7.54$.

\begin{table}[!t]
\centering
\caption{WFA results per fold on forward windows.}
\label{tab:wfa_fold_results}
\begin{tabular}{cccccc}
\toprule
Fold & $SR_{test,i}$ & $\mathrm{MDD}_{test,i}$ & $\mathrm{Calmar}_{test,i}$ & Trades & Gate \\
\midrule
1 & 3.81 & 3.61\% & 3.39 & 258 & \textsc{PASS} \\
2 & 1.36 & 2.89\% & 2.34 & 281 & \textsc{FAIL} \\
3 & 6.20 & 2.30\% & 16.89 & 318 & \textsc{PASS} \\
\midrule
\textit{Mean} & \textit{3.79} & \textit{2.93\%} & \textit{7.54} & \textit{286} & -- \\
\bottomrule
\end{tabular}
\end{table}

\paragraph{Gate verdict.}
Step~2 \textsc{FAIL} (Sharpe $<2.0$), Step~1 and Step~3 \textsc{PASS}. With $q=2/3$, WFA concludes \textsc{PASS}; no fold triggers catastrophic veto ($\mathrm{MDD}<7\%$).

\paragraph{Audit note.}
SR on $W_1^{train}$ is strongly negative while $W_1^{test}$ is high, possibly due to state normalization or train/test simulation-configuration differences. Per the diagnostic-only principle (Section~\ref{sec:stage_wfa}), train metrics do not affect the verdict.

\subsubsection{Stage III (OOS): One-Year Holdout and Verdict}
On the OOS holdout (01/01/2025--31/12/2025) with $\theta^\star$ locked (Section~\ref{sec:stage_wfa}), the strategy achieved:
Sharpe $=2.34$, Calmar $=3.01$, MaxDD $=4.21\%$, $N=1374$ trades.
Per $\mathbf{b}$: Sharpe/Calmar/MaxDD pass. The trades/day criterion requires further cross-referencing.

\subsection{Results Analysis}
\label{sec:analysis}

\subsubsection{Verdict Summary by Decision Gates (IS$\to$WFA$\to$OOS)}
Per the protocol in Section~\ref{sec:strategy_model}, decisions are anchored to \textbf{pre-committed} \textit{decision gates}. Under the current benchmarks (Sharpe $\ge 2.0$, Calmar $\ge 1.5$, MaxDD$<7\%$, and minimum trade density), the case study shows:
(i) Stage I (IS) meets performance thresholds with sufficient trade count,
(ii) Stage II (WFA) achieves majority-pass on forward windows despite one fold failing on Sharpe,
and (iii) Stage III (OOS) recovers benchmark compliance under strict holdout (no tuning).
Therefore, \emph{within the current simulation assumptions}, the alpha may be regarded as \textbf{statistically gate-compliant through time}. However, a ``deploy'' conclusion is valid only when execution assumptions are further verified via the stress envelope (see Section~\ref{sec:analysis_exec_limits}).

\subsubsection{Stage I (IS): Viability and Sample Thickness}
IS results (SR $=2.12$, Calmar $\approx 1.69$, MaxDD $=6.46\%$, $N=2625$ trades) demonstrate that the strategy passes the minimum viability condition and is not based on a handful of observations (a common backtest-overfitting failure mode). In the framework's logic (Section~\ref{sec:stage_is}), this is a necessary condition for advancing from ``alpha idea'' to ``testable candidate.''

An important interpretive note: IS here is used to \emph{lock the candidate space} (plateau/shortlist) rather than to demonstrate generalizability. Accordingly, in the experimental section, IS should be read as evidence that the strategy \textbf{has a sufficiently strong signal} under the stated cost/fill model, not as deployment evidence.

\subsubsection{Stage II (WFA): Sequential Adaptability and Regime Heterogeneity}
WFA forward results show high average performance (SR $=3.79$, MaxDD $=2.93\%$, Calmar $=7.54$), but fold-level decomposition reveals heterogeneity: Step~2 has forward SR $=1.36$ (below the Sharpe threshold), while Step~1 and Step~3 achieve SR $=3.81$ and $=6.20$, respectively. Under the majority-pass design (Section~\ref{sec:stage_wfa}), this case is interpreted as follows:

\begin{itemize}
  \item \textbf{The signal is not absolutely stable across all time segments,} but exhibits \textbf{majority-window robustness} (2/3 folds pass). This represents ``meeting the minimum threshold for advancement'' rather than ``perfect stability.''
  \item \textbf{Step~2 serves as a regime-sensitivity indicator:} this fold should be viewed as a natural \textit{stress-by-time} (not cost stress), suggesting that the alpha may degrade under certain volatility/microstructure regimes.
  \item \textbf{No catastrophic veto:} all forward windows maintain MaxDD below $7\%$, hence no integrity-constraint veto is triggered. This reinforces the framework's argument that \textit{tail-risk control} must be prioritized alongside performance.
\end{itemize}

\paragraph{Audit flag: negative train SR at Step~1.}
As noted in Section~\ref{sec:exp_results}, strongly negative SR on $W_1^{train}$ ($-4.81$) while $W_1^{test}$ is high warrants audit: (i)~re-optimization mechanism, (ii)~state normalization at the test boundary, and (iii)~consistency in Sharpe definition between train/test. Recommendation: report \textbf{fold-level forward metrics + gate} as primary results; train metrics are for audit purposes.

\subsubsection{Stage III (OOS): Strict Holdout and Reasonable Degradation}
OOS results (SR $=2.34$, Calmar $=3.01$, MaxDD $=4.21\%$, $N=1374$ trades) meet the benchmark under strict no-tuning holdout, consistent with the selection-bias reduction objective.

Relative comparison through time:
\begin{itemize}
  \item \textbf{OOS SR vs.\ IS:} $2.34$ vs.\ $2.12$ shows no degradation, suggesting that performance is not merely a peak IS optimum.
  \item \textbf{OOS SR vs.\ WFA (forward mean):} $2.34$ is lower than $3.79$, indicating \textit{performance normalization} upon entering the holdout. This is a commonly observed pattern consistent with the ``not maximizing peak backtest'' thesis.
  \item \textbf{Drawdown integrity:} OOS MaxDD ($4.21\%$) falls between IS ($6.46\%$) and WFA ($2.93\%$), implying that risk does not spike upon entering the holdout.
\end{itemize}
Overall, OOS results provide the strongest evidence in this paper for \textit{generalization under chronology}, but must still be placed in the context of execution assumptions (see below).

\subsubsection{Execution Limitations and Missing Stress Envelope}
\label{sec:analysis_exec_limits}
All current results are produced under \textit{ideal execution} assumptions (no latency, no adverse slippage beyond the MT5 tick model). Since the paper highlights Gap~C, the absence of a stress envelope implies:
\begin{itemize}
  \item \textbf{Time-robustness under the current cost model can be established,}
  \item but \textbf{execution-robustness under microstructure/cost inflation cannot yet be concluded.}
\end{itemize}
In the absence of time to re-run multiple batteries, the scientifically appropriate presentation is:
(i) explicitly state this as \textbf{baseline validation} with stress-testing as \textbf{planned work},
(ii) retain the stress protocol as a \textit{template} (without presenting results),
(iii) add a \textbf{scope limitation} statement at the end of the experimental section: ``deployment-readiness requires passing a cost/slippage stress envelope; results herein should be interpreted under baseline MT5 assumptions.''

\subsubsection{Implications for Future Alpha Evaluation and Ranking}
The WFA/OOS portion of the case study should be understood as a \textbf{single validation unit} for \emph{one} alpha. For future multi-alpha leaderboards, the framework suggests a final ranking principle that is \textit{post-validation} in nature: rank only alphas that have \textsc{PASS}ed IS$\to$WFA$\to$OOS (and eventually the stress envelope), then compare using a pre-committed utility (e.g., prioritize MaxDD/Calmar over SR when the mandate is \textit{capital preservation}). This clearly separates the \textbf{deployment decision protocol} from the \textbf{research portfolio report} (portfolio/roadmap), consistent with the spirit of the introduction.


\section{Post-Validation Comparison of Alpha Variants}
\label{sec:alpha_comparison}

This section constitutes a \textit{post-validation report}, \textbf{not} a step within the framework's decision gates. The objective is to compare four alpha variants that have \textsc{pass}ed the IS$\to$WFA$\to$OOS pipeline under the same data setup and execution assumptions described in Section~\ref{sec:exp_setting}, serving research orientation and capital allocation (portfolio view).

\subsection{Comparison Protocol}
\label{sec:comparison_protocol}

Four candidates (\texttt{v1}--\texttt{v4}) are evaluated according to three principles:
\begin{enumerate}
    \item \textbf{OOS-first ranking:} primary ranking is based on OOS holdout (Sharpe, Calmar, $\mathrm{MDD}_{eq}$), as this constitutes the strictest ``blind'' test~\cite{pardo,bailey14}.
    \item \textbf{WFA dispersion:} $\overline{SR}_{\mathrm{WFA}}$ and inter-fold range measure stability across market regimes (diagnostic, not used for ranking).
    \item \textbf{Diagnostic-only:} train metrics and resilience ratio $\eta_i$ serve analysis only; they are not used to modify rankings or parameters after $\theta^\star$ has been locked (Section~\ref{sec:stage_wfa}).
\end{enumerate}

\subsection{Stage-by-Stage Results}
\label{sec:alpha_results}

Table~\ref{tab:alpha_summary} summarizes Sharpe, Calmar, and $\mathrm{MDD}_{eq}$ across the three stages. \textbf{Bold} values indicate the leading candidate at OOS for each metric.

\begin{table}[t]
\centering
\caption{Performance of four alphas across IS\,/\,WFA\,/\,OOS.}
\label{tab:alpha_summary}
\resizebox{\linewidth}{!}{%
\begin{tabular}{l ccc ccc ccc c}
\toprule
 & \multicolumn{3}{c}{Sharpe} & \multicolumn{3}{c}{Calmar} & \multicolumn{3}{c}{MaxDD (\%)} & Trades \\
\cmidrule(lr){2-4} \cmidrule(lr){5-7} \cmidrule(lr){8-10} \cmidrule(lr){11-11}
Alpha & IS & WFA & \textbf{OOS} & IS & WFA & \textbf{OOS} & IS & WFA & \textbf{OOS} & (OOS) \\
\midrule
\texttt{v1} & 2.43 & 3.18 & 2.19 & 1.85 & 6.17 & 2.74 & 6.96 & 3.04 & 4.43 & 1\,375 \\
\texttt{v2} & 2.29 & 2.93 & 2.56 & 1.83 & 5.39 & \textbf{3.52} & 6.52 & 3.30 & 4.41 & 1\,428 \\
\texttt{v3} & 2.20 & 2.91 & \textbf{2.61} & 1.72 & 5.39 & 3.48 & 6.69 & 3.35 & 4.36 & 1\,427 \\
\texttt{v4} & 2.12 & 3.79 & 2.34 & 1.69 & 7.54 & 3.01 & 6.46 & 2.93 & \textbf{4.21} & 1\,374 \\
\bottomrule
\end{tabular}}
\end{table}

\paragraph{Key findings (OOS holdout).}
\begin{itemize}
    \item \textbf{Sharpe:} \texttt{v3} leads (2.61), followed by \texttt{v2} (2.56); the difference is small and no statistical significance test has been conducted.
    \item \textbf{Calmar:} \texttt{v2} is highest (3.52). If a deployment threshold of $\mathrm{Calmar}_{\mathrm{OOS}}>3.0$ is applied, \texttt{v1} (2.74) does not qualify.
    \item \textbf{MaxDD:} \texttt{v4} is lowest (4.21\,\%), demonstrating superior tail-risk control.
    \item \textbf{Control observation:} \texttt{v1} has the highest IS Sharpe but the weakest OOS---illustrating the risk of selecting alphas by IS peak~\cite{bailey14,dsr14}.
\end{itemize}

\subsection{WFA Stability Diagnostics}
\label{sec:wfa_dispersion}

Table~\ref{tab:wfa_steps} presents $SR_{test,i}$ across each forward fold along with the mean and range.

\begin{table}[t]
\centering
\caption{Forward Sharpe by fold (WFA 3-fold).}
\label{tab:wfa_steps}
\begin{tabular}{l ccc cc}
\toprule
 & \multicolumn{3}{c}{$SR_{test,i}$} & & \\
\cmidrule(lr){2-4}
Alpha & Fold\,1 & Fold\,2 & Fold\,3 & $\overline{SR}_{\mathrm{WFA}}$ & Range \\
\midrule
\texttt{v1} & 2.97 & 1.37 & 5.19 & 3.18 & 3.82 \\
\texttt{v2} & 2.56 & 1.14 & 5.08 & 2.93 & 3.94 \\
\texttt{v3} & 2.53 & 1.17 & 5.02 & 2.91 & 3.85 \\
\texttt{v4} & 3.81 & 1.36 & 6.20 & 3.79 & 4.84 \\
\bottomrule
\end{tabular}
\end{table}

All four alphas are relatively weak in Fold\,2 ($SR\approx 1.1$--$1.4$) and strong in Fold\,3 ($SR\approx 5$--$6$), indicating a common regime dependence---this is precisely why rolling WFA is needed rather than relying on a single backtest (see Section~\ref{sec:stage_wfa}).

\subsection{Equity-Path Diagnostics (OOS)}
\label{sec:oos_equity_diagnostics}

\begin{figure}[!t]
    \centering
    \includegraphics[width=0.95\linewidth]{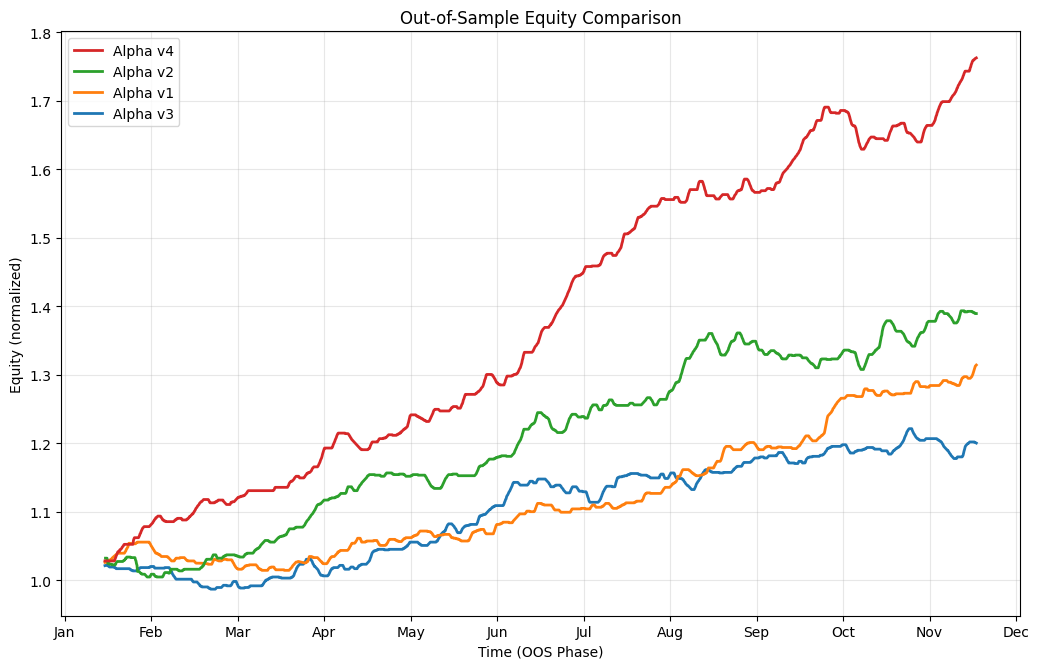}
    \caption{Normalized OOS equity (\texttt{v1}--\texttt{v4}).}
    \label{fig:oos_equity_oos}
\end{figure}

Figure~\ref{fig:oos_equity_oos} enables visual inspection of failure modes that aggregate statistics may obscure: (i)~profit concentrated in a short segment, (ii)~extended plateaus, or (iii)~slow recovery after drawdown.

\subsection{Trade-Off and Mandate-Dependent Recommendations}
\label{sec:mandate_tradeoff}

Ranking depends on the objective function:
\begin{itemize}
    \item \textbf{Maximizing risk-adjusted return:} \texttt{v3} (Sharpe) or \texttt{v2} (Calmar) are the leading candidates.
    \item \textbf{Capital preservation (tail-risk):} \texttt{v4} has the lowest $\mathrm{MDD}_{\mathrm{OOS}}$; the trade-off is lower Sharpe/Calmar compared to \texttt{v2/v3}.
\end{itemize}
The rank reversal when switching from maximizing Sharpe to minimizing MaxDD reflects the fundamental trade-off between performance and tail-risk control~\cite{dsr14}, underscoring that \emph{no single metric suffices} for alpha ranking---a multi-dimensional criterion set tied to a specific mandate is required.

\subsection{Limitations of the Post-Validation Report}
\label{sec:comparison_limits}

\begin{itemize}
    \item Results are based on \textit{ideal execution} assumptions (Section~\ref{sec:dataset_exec}); rankings may change under a cost/slippage stress envelope.
    \item Comparing four variants from the same search process increases degrees of freedom at the research level; interpreting small differences between \texttt{v2} and \texttt{v3} requires supplementary tests (Deflated Sharpe~\cite{dsr14}, SPA~\cite{hansen05}, or HAC confidence intervals~\cite{neweywest}).
\end{itemize}


\section{Discussion and Conclusion}
\label{sec:conclusion_vn}

\subsection{Practical Implications for Deployment and Governance}
This paper treats \textbf{deployment} as an auditable decision rather than a backtest optimization outcome. Accordingly, the primary ``deliverable'' is not a single alpha but a \textbf{standardized validation protocol} that produces a \textsc{PASS}/\textsc{FAIL} verdict following IS$\to$WFA$\to$OOS chronology and \textbf{pre-committed} \textit{decision gates}.

\paragraph{Minimum deployment checklist.}
A strategy is deemed eligible for advancement only if it simultaneously satisfies:
\begin{itemize}
    \item \textbf{Chronology \& no-peeking:} $\theta^\star$ locked after WFA; no tuning upon opening OOS (Section~\ref{sec:stage_wfa}).
    \item \textbf{Restricted DoF:} WFA re-selects within $\Omega_{\mathrm{stable}}$/shortlist; $\Theta$ is not re-opened after Stage~I (Section~\ref{sec:stage_is}).
    \item \textbf{Gate on forward windows:} majority-pass + catastrophic veto (Section~\ref{sec:stage_wfa}, Algorithm~\ref{alg:wfa}).
    \item \textbf{Execution-aware reporting:} execution assumptions are explicitly declared; results are interpreted within that context rather than defaulting to ``deployable.''
\end{itemize}

\paragraph{Governance/audit: requirements for auditable research.}
To reduce selection bias and enhance reproducibility, research should be accompanied by a minimum ``evidence pack'':
\begin{itemize}
    \item \textbf{Search transparency:} description of search budget, grid/trial size, permitted parameter dimensions, and locking timing.
    \item \textbf{Artifact logging:} backtest configuration, data version, seed (if applicable), and stage-by-stage results (IS/WFA/OOS) under a common metric standard.
    \item \textbf{Decision trace:} explicit pass/fail rules (benchmark $\mathbf{b}$, majority threshold $q$, catastrophic veto) and failure-mode-to-verdict mapping (\textbf{Reject} vs.\ \textbf{Refactor}).
\end{itemize}
These requirements separate ``attractive results'' from ``valid process,'' making conclusions falsifiable: if OOS is opened and the gate fails, the strategy is eliminated without ``rescue tuning.''

\subsection{Limitations}
Results should be interpreted in the context of the following limitations (see also Section~\ref{sec:analysis_exec_limits}):
\begin{itemize}
    \item \textbf{Simplified execution:} ideal execution assumed; latency/slippage not explicitly modeled---increases optimism risk for microstructure-sensitive strategies.
    \item \textbf{Single asset/broker:} USDJPY M5 case study on MT5/Exness; transferability has not been verified.
    \item \textbf{Incomplete stress/ablation:} stress envelope described as a template; current results constitute baseline validation.
    \item \textbf{Multi-alpha selection bias:} comparing v1--v4 increases DoF at the research level; small differences require supplementary testing~\cite{dsr14,hansen05}.
\end{itemize}

\subsection{Future Directions}
Direct extensions to enhance deployability and scientific rigor include:
\begin{itemize}
    \item \textbf{Multi-asset \& multi-timescale validation:} replicate the protocol across multiple FX pairs/assets and timeframes; report regime-conditional stability (trend/range, volatility regimes).
    \item \textbf{More realistic execution models:} incorporate explicit latency/slippage models, cost-ladder stress envelopes (spread/commission/slippage inflation), and ``point-of-failure'' testing.
    \item \textbf{Portfolio-level governance:} transition from single-alpha evaluation to portfolio selection/retention/exit processes (correlation, tail co-movement, risk budgeting), with post-deployment monitoring rules.
    \item \textbf{Pre-registration and selection-bias testing:} pre-commit benchmarks, number of trials, and alpha selection criteria; apply appropriate tests/corrections for multiple hypothesis testing.
    \item \textbf{Live paper trading:} add a paper/live-sim stage as a ``deployment rehearsal'' to verify microstructure assumptions before real capital allocation.
\end{itemize}

\subsection{Conclusion}
This paper proposes the \textbf{AlgoXpert Alpha Research Framework}, a decision-oriented IS--WFA--OOS protocol designed to reduce overfitting and improve robustness when transitioning from backtest to operations. The core contributions of the framework are (i) parameter selection via \textbf{stability regions} rather than extreme points, (ii) \textbf{purged rolling WFA} to mitigate leakage for stateful/path-dependent strategies, (iii) integrated \textbf{defense-in-depth safeguards} following an execution-aware approach, and (iv) \textbf{auditable pass/fail decision gates}.

Within the current illustrative scope, the framework is \textbf{validated} in the sense that the chronological validation procedure produces consistent and traceable decisions, while surfacing failure modes requiring audit (e.g., train--test discrepancies by fold) rather than concealing them behind aggregate statistics. Conversely, the framework \textbf{should not} be interpreted as a profit guarantee; simulation assumptions (particularly latency/slippage) and the completeness of stress testing/ablation are boundary conditions for deployability inference.

In summary, the primary value of the framework lies in standardizing the \textbf{deployment decision process} rather than maximizing ``peak backtest'': an alpha advances only upon passing pre-committed decision gates, under declared execution assumptions, with degrees of freedom controlled to reduce selection bias.


\end{document}